\documentclass{article}

\usepackage{PRIMEarxiv}

\usepackage[utf8]{inputenc} 
\usepackage[T1]{fontenc}    
\usepackage{hyperref}       
\usepackage{url}            
\usepackage{booktabs}       
\usepackage{amsfonts}       
\usepackage{nicefrac}       
\usepackage{microtype}      
\usepackage{lipsum}
\usepackage{fancyhdr}       
\usepackage{graphicx}       
\graphicspath{{media/}}     
\usepackage{amsmath}
\usepackage{rotating}
\usepackage{hyperref}
\hypersetup{
    colorlinks=true,
    urlcolor=red,
    }

\title{A blended distance to define "people-like-me"}

\author{
  Anaïs Fopma, Mingyang Cai, Stef van Buuren \& Gerko Vink \\
  Department of Methodology \& Statistics \\
  Utrecht University \\
  Utrecht, the Netherlands\\
}

\begin{document}
\maketitle

\begin{abstract}
Curve matching is a prediction technique that relies on predictive mean matching, which matches donors that are most similar to a target based on the predictive distance. Even though this approach leads to high prediction accuracy, the predictive distance may make matches look unconvincing, as the profiles of the matched donors can substantially differ from the profile of the target. To counterbalance this, similarity between the curves of the donors and the target can be taken into account by combining the predictive distance with the Mahalanobis distance into a `blended distance' measure. The properties of this measure are evaluated in two simulation studies. Simulation study I evaluates the performance of the blended distance under different data-generating conditions. The results show that blending towards the Mahalanobis distance leads to worse performance in terms of bias, coverage, and predictive power. Simulation study II evaluates the blended metric in a setting where a single value is imputed. The results show that a property of blending is the bias-variance trade off. Giving more weight to the Mahalanobis distance leads to less variance in the imputations, but less accuracy as well. The main conclusion is that the high prediction accuracy achieved with the predictive distance necessitates the variability in the profiles of donors.
\end{abstract}


\section{Introduction}
The first three years of childhood form a crucial stage in determining children’s subsequent development and health outcomes.\cite{straatmann_how_2018} For this reason, growth monitoring is considered to be an integral part of paediatrics. It can aid in the identification of problems in development such as growth stunting, and ensure timely treatment or intervention to improve the child’s health.\cite{cordeiro_childs_2019} However, growth monitoring solely provides insights in the past and current developmental stages of the child. Growth curve modeling, on the other hand, can be used to predict future development. It could therefore provide more specific answers to questions health professionals, parents, and insurance companies may have, such as: ‘Given what I know of the child, how will it develop in the future?’ and ‘Does this child get the most effective treatment  available?’\cite{van_buuren_curve_2014} 

\subsection{Curve matching}
An approach currently used for growth curve modeling is curve matching. Curve matching\cite{van_buuren_curve_2014} is a nearest neighbour technique for individual prediction that constructs a prediction by aggregating the histories of “people-like-me”. It aims to predict the growth of a target child by using the data of other children that are most similar to the target child. 

In order to select these donors, some form of similarity needs to be defined to match the donors to the target child. Therefore, the key question is: How are good matches obtained? The current approach uses predictive mean matching (PMM). PMM is a multiple imputation technique that makes use of an existing donor database, containing the growth data of children who are older than the target child. Therefore, the information of these children at a later age is available. The first step is to fit a linear regression model on the donor database. Then, this model is used to predict the values for all donors and for the target at a certain point in the future, for example at 14 months. Finally, the distance between the predicted value of each of the donors and the predicted value of the target is calculated, which is referred to as the predictive distance. A number of donors – usually five - with the smallest predictive distance are selected as the best matches. Their growth curves are then plotted and point estimates can be calculated by averaging the measurements. The growth patterns of the matched children thus suggest how the target child might develop in the future. 

\subsection{Alternative approach}
PMM has proven to be promising in growth curve matching and the advantage of this technique is its high prediction accuracy.\cite{van_buuren_curve_2014} However, there are two reasons to move beyond the predictive distance used in PMM and investigate an alternative metric. Firstly, PMM requires users of curve matching to select a particular future time point to base the matches on (e.g. 14 months of age). In some cases, it may be difficult to choose this time point, especially when the ‘future’ is more vaguely defined as a time interval.\cite{van_buuren_broken_2020} Secondly, the predictive distance may make the matches look unconvincing. The trajectories of the selected donors may all be close to the prediction for the target child at 14 months, but this does not imply that the histories are identical. After all, different profiles may lead to the same predicted value. Consequently, the curves of some of the matches may be quite far from the curve of the target child. Some users of curve matching feel that such discrepancies are undesirable, as these matches do not appear to be \textit{people-like-me}.\cite{van_buuren_broken_2020} It is useful to investigate these shortcomings not only for improving growth prediction but also for other applications of multiple imputation, such as patient recovery after an operation, prediction of longevity, and decision-making when more than one treatment is available. \cite{van_buuren_curve_2014}

For the aforementioned reasons, the practical implementation and use of curve matching can in theory be improved by combining the predictive distance with another distance measure, thus creating a “blended distance” measure. Such a blended metric would take into account historical similarity between the donors and the target. For example, when blending the predictive distance with the Mahalanobis distance, more weight is given to similarities between units in the full predictor space. This would theoretically lead to the selection of donors with profiles more similar to the target, and therefore to the selection of true people-like-me. The objective of this study is to implement such a blended distance measure and to investigate its properties, blend ratio, and the validity of its resulting inferences.

\section{Methods}
\subsection{Blended metric}
The blended distance measure in this manuscript is a weighted version of the predictive distance (PD) and the Mahalanobis distance (MD). The PD is the distance between the predicted value of a donor and the predicted value of the target at a particular future time point. The MD is defined as the distance between two $N$ dimensional points scaled by the variation in each component of the point. For example, if $\vec{x}$ and $\vec{y}$ are two points from the same distribution with covariance matrix ${\rm  C}$, then the  MD is given by
    \begin{equation}
    ((\vec{x} - \vec{y})' {\rm  C}^{-1} (\vec{x} - \vec{y}) )^{\frac{1}{2}}.
    \end{equation}
Two potential versions of the blended metric will be compared: one that uses ranking and one that uses scaling. In theory, these two versions of the blended distance should yield similar results. However, the scaled version would be computationally more convenient as no rank-orders need to be computed. Both versions are evaluated to study their performance.
For the ranked blended distance (RBD), the PD and the MD are first calculated for each donor. Then, the \textit{k} donors with the lowest values for both the PD and the MD are selected. In order to do so, the rank is calculated for the PD and the MD, where ties are randomly broken. The RBD is given by:
    \begin{equation}
    \textit{RBD} = p \cdot \textrm{rank}_\textit{PD} + (1-p) \cdot \textrm{rank}_\textit{MD},
    \end{equation}
where $\textrm{rank}_{PD}$ is the rank for the PD, $\textrm{rank}_\textit{MD}$ is the rank for the MD, $p$ is the blending factor (or weight) assigned to $\textrm{rank}_\textit{PD}$, and $0\leq p \leq 1$. The \textit{k} donors with the lowest values on the RBD are selected as the best matches.

The scaled blended distance (SBD) is created similarly, but both the PD and the MD are scaled before combining them. The SBD is given by:
    \begin{equation}
    \textit{SBD} = p \cdot \frac{\textit{PD} - \bar{x}_\textit{PD}}{\sigma_\textit{PD}} + 
    (1-p) \cdot \frac{\textit{MD} - \bar{x}_\textit{MD}}{\sigma_\textit{MD}},
    \end{equation}
where $\bar{x}_\textit{PD}$ is the mean of the PDs, $\sigma_\textit{PD}$ their standard deviation, $\bar{x}_\textit{MD}$ is the mean of the MDs, and $\sigma_\textit{MD}$ their standard deviation.

\begin{figure}[t]
\centerline{\includegraphics[width=\textwidth,height=\textheight,keepaspectratio]{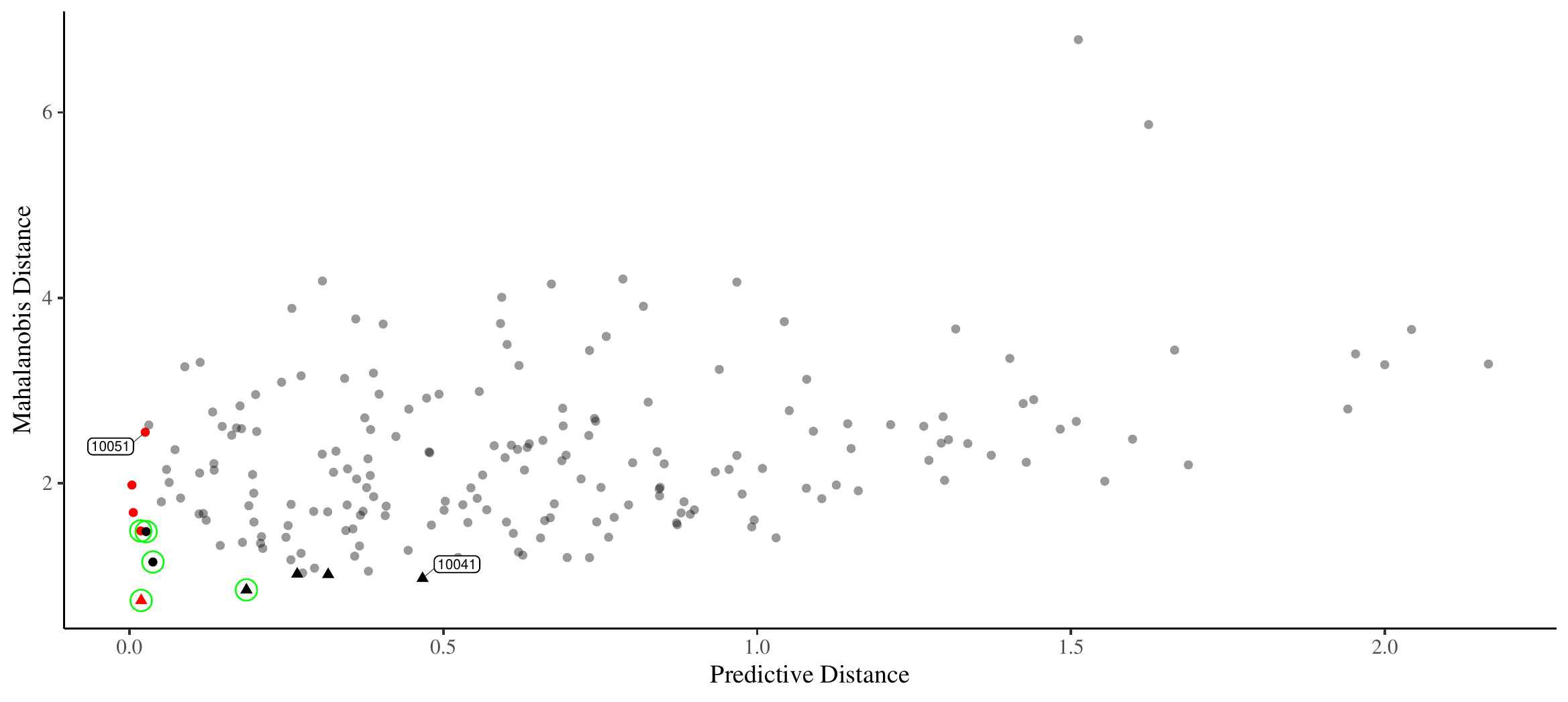}}
\caption{MD plotted against PD for each of the 199 donors. The donors in red are the five matches with the smallest PD, the triangular donors those with the smallest MD, and the donors circled in green those with the smallest blended distance.\label{fig1}}
\end{figure}

As an example, the blended distance is illustrated in Figure \ref{fig1}. Here, the data of 200 children from the \textit{Sociaal Medisch Onderzoek Consultatiebureau Kinderen} (SMOCK) study are used.\cite{herngreen_growth_1994} The first subject is taken as the target, the 199 other subjects as the donors. For all donors, the MD for the measurements during the first six months of growth is calculated. In addition, the PD between each donor and the target is calculated. In the figure, the MD and PD are plotted against each other. The red donors are the five matches with the smallest PD, where especially subject 10051 has a large MD. The triangular donors are the five matches with the smallest MD, where especially subject 10041 has a large PD. A weighted blended metric would balance the two distance measures, such that the donors with a low value for both distance measures are chosen. These are circled in green. 

\subsection{Simulations}
In order to investigate the properties of the blended distance measure, two simulation studies are conducted. Each study is described in accordance with the ADEMP-structure for reporting simulation research.\cite{morris_using_2019} The aims of each study, the different versions of the blended metric (i.e. the methods), the data-generating mechanisms, and the estimand and performance measures are discussed.

\subsection{Software}
R version 4.2.0 (2022-04-01)\cite{R2022} is used to simulate the data and perform the analyses. The \texttt{mice.impute.pmm()} function in the \texttt{mice}\cite{van_buuren_mice_2011} package is used to perform PMM and an \href{https://github.com/anaisfopma/mice/tree/development}{adaptation of this function} is used to calculate the blended distance. Instructions and scripts to reproduce the simulation results are available in the \href{https://github.com/anaisfopma/Blended-Distance-Thesis}{research archive} of this project.

\section{Simulation study I}

\subsection{Aims}
The main objective of this simulation study is to investigate what the properties of the blended metric are. More specifically, I want to answer the following questions:
\begin{enumerate}
    \item Does performance differ when donors are selected with a ranked as opposed to a scaled blended distance?
    \item Does a blending factor of 1 yield results identical to those obtained by PMM (i.e. based on the PD only)?
    \item How is the performance of the blended metric related to the missingness mechanism, the proportion of missingness in the data, and the distribution of the data?
    \item Do the conclusions under 3. differ for varying levels of correlation in the predictor space? 
    \item Is there a penalty from blending in terms of reduced predictability? 
\end{enumerate}

I expect that the ranked and scaled versions of the blended metric yield similar results, as donors with a high ranking will likely have a small value on the scaled corresponding distance, and vice versa. Furthermore, I expect that blending with a factor of 1 does indeed give the same results as PMM does, as this entails that full weight is given to the PD. As pointed out before, PMM has been shown to result in high prediction accuracy. Therefore, I expect that the predictability of the blended distance will decrease as the blending factor favours the MD. When the correlation in the data is low, the prediction model will fit poorly and I expect the blended metric to perform worse when more weight is given to the PD. When the correlation in the data is high, the prediction model will fit better, and the prediction model will explain more variance in the outcome. In this case, I expect the blended metric to perform better when more weight is given to the PD. Finally, I expect that the blended metric will perform worse in skewed data when more weight is given to the MD.

\subsection{Metrics}
For the blended metric, blending factors of respectively 1, 0.5, and 0 are evaluated. A blending factor of 1 implies that the blended distance is equal to the PD, whereas a weight of 0 implies that it is equal to the MD. Therefore, a blending factor of 0.5 gives equal weight to both distance measures. PMM will be used as a reference in order to evaluate whether it does indeed obtain the same results as the blended metric with a blending factor of 1. Using PMM, both the ranking and scaling methods with three different blending factors each, this results in seven different versions of the blended metric that are evaluated. 

\subsection{Data-generating mechanisms}
In order to answer the previous questions, the blended distance measures are evaluated in simulated data that meet different conditions. All data are generated from one of 24 data-generating mechanisms, with equal means, but with varying missingness proportions, missingness mechanisms, distributions, and variance-covariance matrices. I explain the process of obtaining these data-generating mechanisms below.

Three continuous predictor variables $X_1$, $X_2$, and $X_3$ are defined and one continuous outcome $Y$ is defined: $Y = X_1 + X_2 + X_3 + \epsilon$, where $\epsilon$ is random noise drawn from a normal distribution with $\mu = 0$ and $\sigma = 7$. The distribution of the data is varied over two conditions. For the first conditions, the data-generating mechanism of the predictor space is a multivariate normal distribution, $X = \mathcal{N}({\mu}, \mathrm{\Sigma})$, with mean vector $\mu = \mathrm{[10,10,10]}$. For the second condition, it is a strongly skewed multivariate distribution, which is obtained by transforming the predictors\cite{vink_predictive_2014}: $X = X^{12}/max\{X^{11}\}$. 

The correlation in the data is varied over three conditions. The covariance matrix $\Sigma$ for the populations is given by: 

$$
  {\Sigma} = \begin{bmatrix}
                    1  & \sigma^{2}\rho & \sigma^{2}\rho  \\
                    & 1 & \sigma^{2}\rho  \\
                    &  & 1 \\
                    \end{bmatrix},
$$

where the off-diagonal elements are 0 for the first simulation condition, 0.1 for the second condition, and 0.7 for the third condition. 

The proportion of missingess in the outcome variable is varied over two conditions. The first condition simulates a setting with 25\% missingness, the second a setting with 50\% missingness.

Finally, the missingness mechanism is varied over two conditions. The first concerns a missing completely at random (MCAR) mechanism, where missingness does not depend on the values of the data, missing or observed. \cite{little_statistical_2019} The second concerns a missing at random (MAR) right mechanism. This means that missingness does depend on the data, but only through observed components of the data. \cite{little_statistical_2019} The MAR right mechanism deletes higher values for positively correlated data, so the distribution of the observed data shifts to the left.\cite{van_buuren_flexible_2018} The MAR right mechanism is considered to be one of the more extreme missingness mechanisms, as it is asymmetrical and causes common statistics such as the mean to be biased.\cite{van_buuren_flexible_2018} It is unlikely to see such a mechanism in practice, but it is helpful in simulation studies to test the limits of a method. If the distance measure performs well under this mechanism, it will also do so in less extreme situations that are more likely to be encountered in practice.\cite{van_buuren_flexible_2018}

I consider a full-factorial simulation study design, where each of the possible combinations of weighting and data-generating mechanisms are evaluated. As there are seven different methods and 24 different data-generating mechanisms, the simulation will yield 168 results. From each data-generating mechanism, a single complete sample of size 500 is drawn. The 1000 simulated incomplete versions of these data (with missingess proportion of 25\% or 50\%) are then imputed. The finite population pooling rules by Vink \& Van Buuren\cite{vink_pooling_2014} are used to obtain inferences over the simulated results. 

\subsection{Estimand and performance measures}
The estimands of interest in this study are the predicted (i.e. imputed) values. I assess the statistical validity of each metric under each combination of conditions by means of the bias, coverage, and proportion of explained variance. 

\subsection{Results}
\begin{figure}[t]
\centerline{\includegraphics[width=\textwidth,height=\textheight,keepaspectratio]{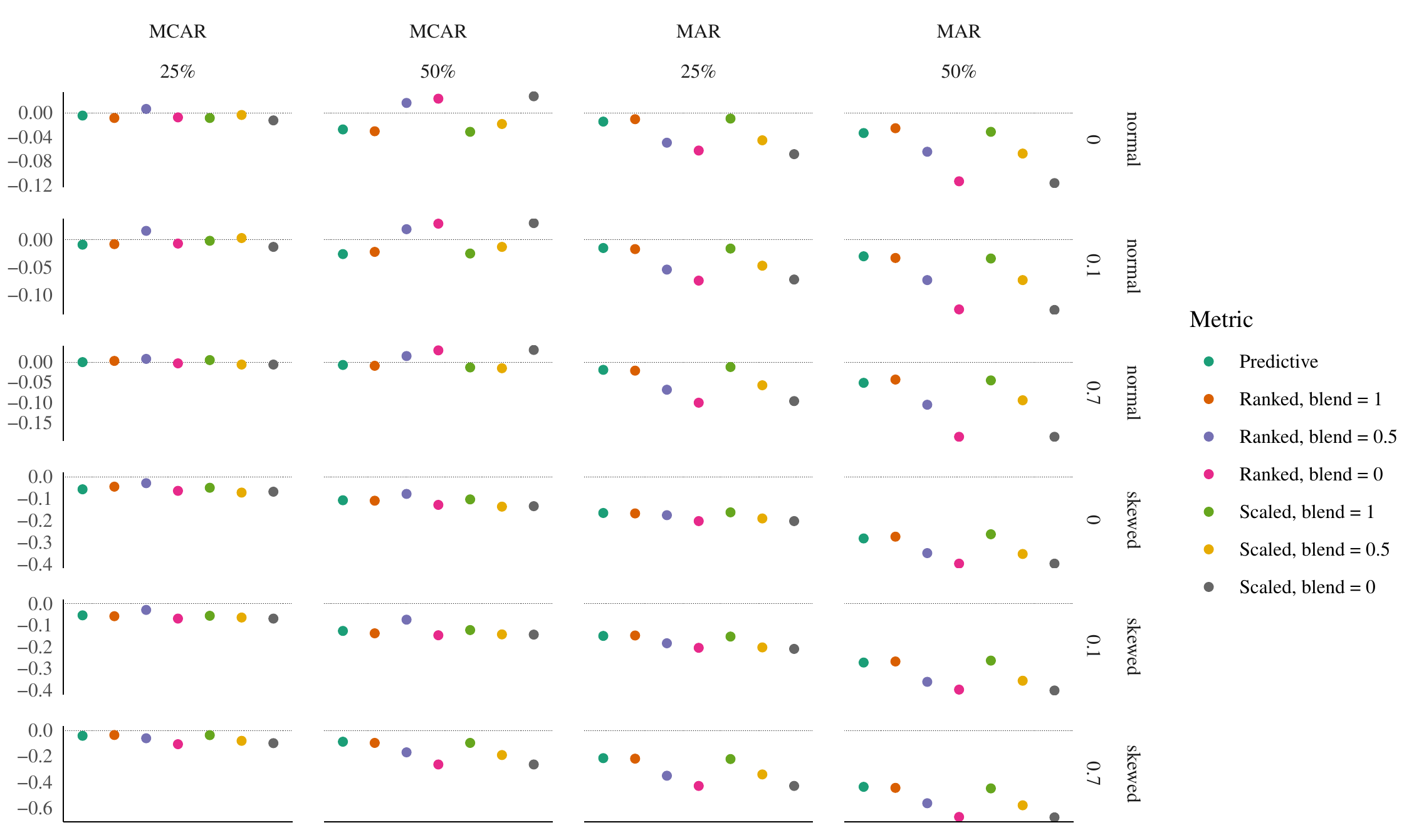}}
\caption{Bias results per condition, where each individual plot shows the results for the seven methods. A reference line is given at bias = 0. Above the plots, the condition combinations of missingness mechanism (MCAR, MAR right) and missingness proportion (25\%, 50\%) are given. On the right, the condition combinations of distribution (normal, skewed) and correlation ($\rho = 0, \rho = 0.1, \rho = 0.7$) \textrm{ are given.}\label{fig2}}
\end{figure}

\begin{figure}[t]
\centerline{\includegraphics[width=\textwidth,height=\textheight,keepaspectratio]{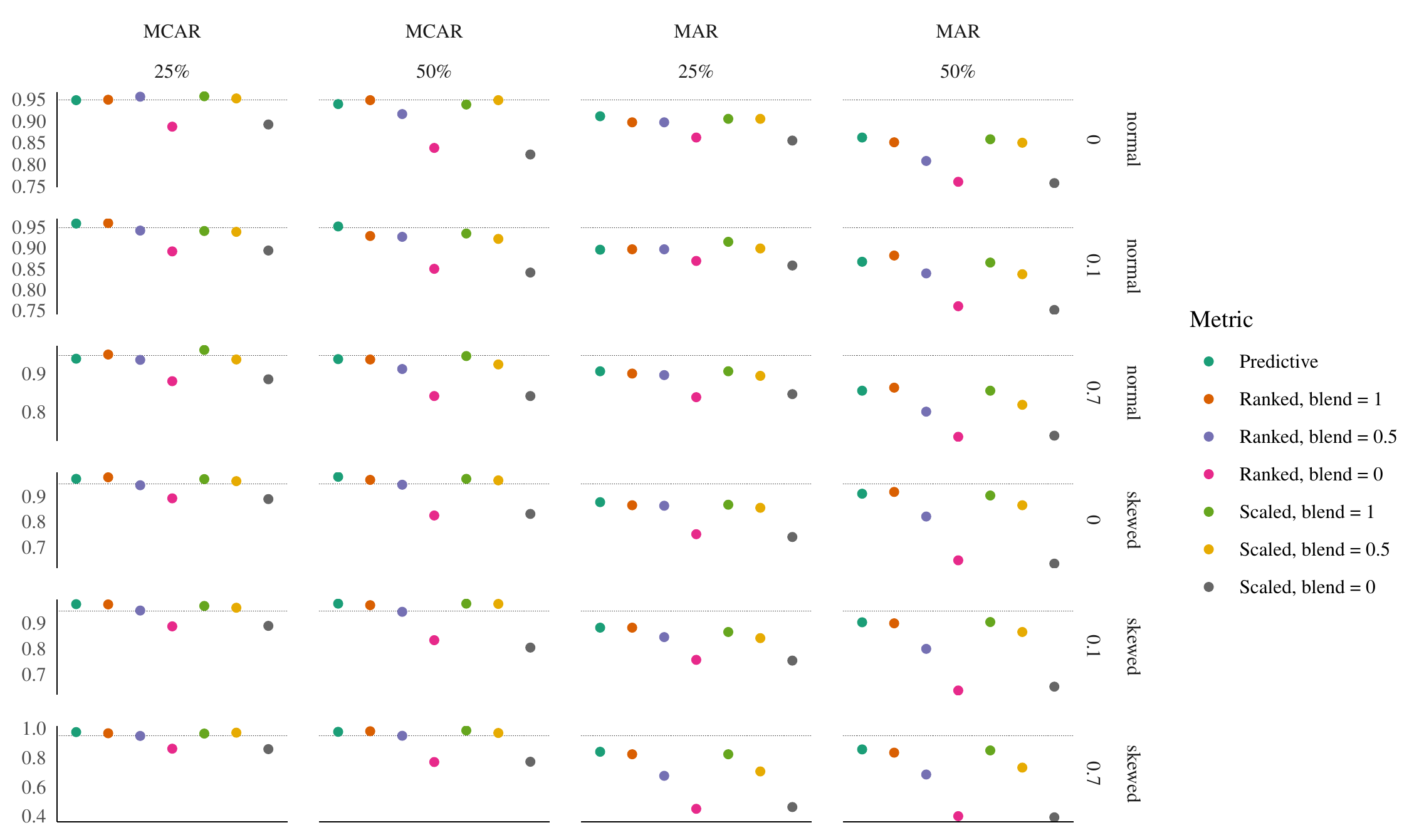}}
\caption{Coverage results per condition, where each individual plot shows the results for the seven methods. A reference line is given at coverage = 0.95. Above the plots, the condition combinations of missingness mechanism (MCAR, MAR right) and missingness proportion (25\%, 50\%) are given. On the right, the condition combinations of distribution (normal, skewed) and correlation ($\rho = 0, \rho = 0.1, \rho = 0.7$)\textrm{ are given.} \label{fig3}}
\end{figure}

\begin{figure}[t]
\centerline{\includegraphics[width=\textwidth,height=\textheight,keepaspectratio]{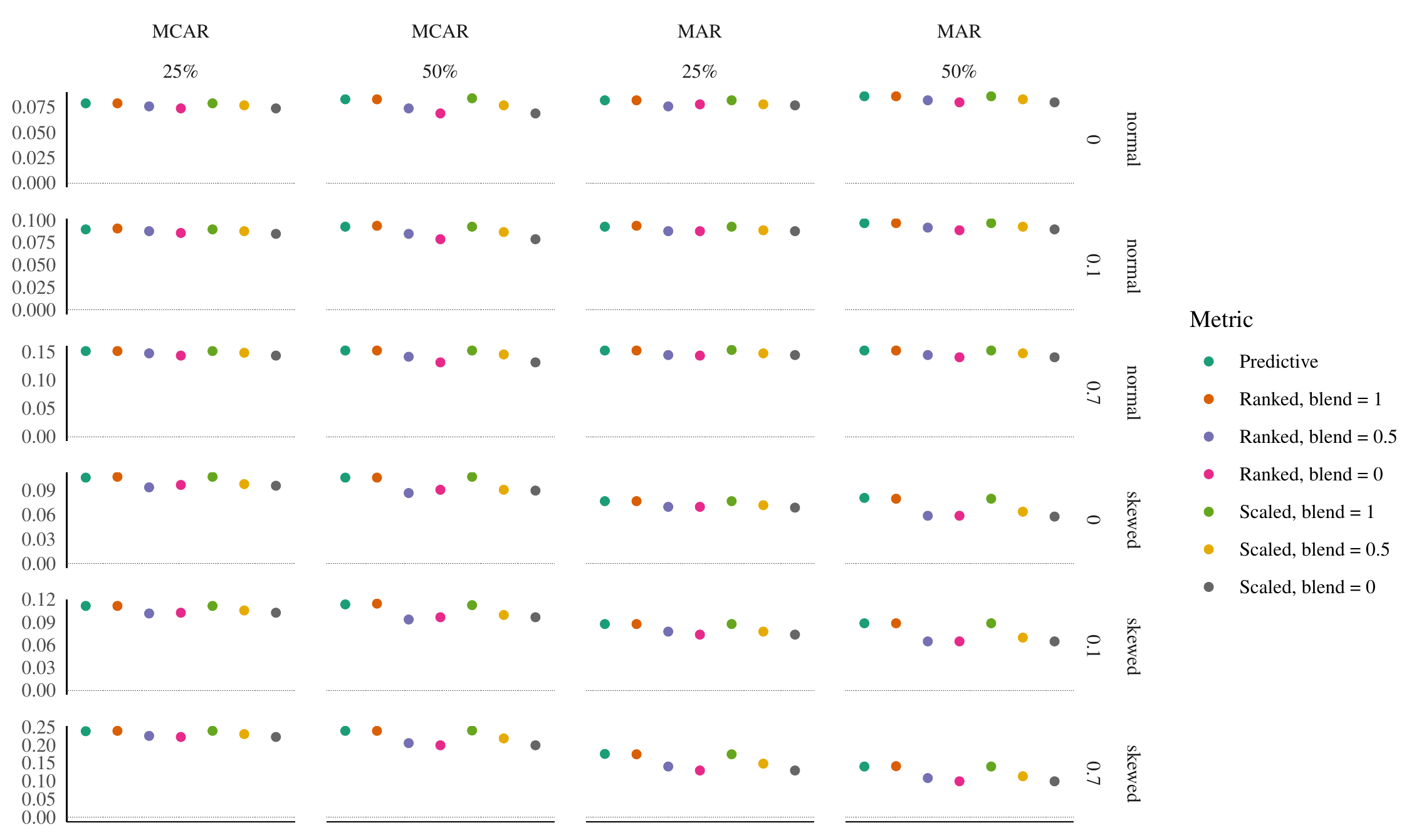}}
\caption{R squared results per condition, where each individual plot shows the results for the seven methods. A reference line is given at 0. Above the plots, the condition combinations of missingness mechanism (MCAR, MAR right) and missingness proportion (25\%, 50\%) are given. On the right, the condition combinations of distribution (normal, skewed) and correlation ($\rho = 0, \rho = 0.1, \rho = 0.7$) \textrm{ are given.}\label{fig4}}
\end{figure}

Table \ref{tab1} through Table \ref{tab7} in Appendix \ref{app1} display the simulation results for each of the seven methods. Each table specifies the data-generating mechanisms in the left columns by indicating the missingness mechanism, missingness proportion, skewness of the distribution, and correlation in the data. Figure \ref{fig2} visualises the results for the coverage, Figure \ref{fig3} those for the bias, and Figure \ref{fig4} those for the explained variance. I discuss the results below on the basis of the research questions. 

\subsubsection{Comparison of ranked and scaled blended distance}
The ranked and scaled versions of the blended distance measure yield similar simulation results. When comparing the results for blending factor = 1 (full PD) in Table \ref{tab2} and Table \ref{tab5}, they show that the SBD yields slightly higher coverages but larger biases overall. When comparing the results for blending factor = 0.5 in Table \ref{tab3} and Table \ref{tab6}, they show that the SBD outperforms the RBD. The results for blending factor = 0 (full MD) in Table \ref{tab4} and Table \ref{tab7} show that the RBD outperforms the SBD. Overall, the RBD performs slightly better, except when the blending factor is set to 0.5. As most of the results are similar, however, the SBD might be preferable, as it is computationally more efficient to use. 

\subsubsection{Comparison of PMM and blending factor = 1}
In both blended metrics, a blending factor of 1 indicates that full weight is given to the PD. Therefore, using a blending factor of 1 should yield results identical to those obtained by PMM. Even though the results in Table \ref{tab1}, Table \ref{tab2} and Table \ref{tab5} are similar, they are not identical. This is likely due to the fact that a matching function written in C is used in the original \texttt{mice.impute.pmm()} function but not in the \texttt{mice.impute.blended()} function. Therefore any differences are likely due to the different underlying code. In some cases, particularly in the MCAR conditions, the both the RBD and SBD with a blending factor of 1 perform slightly better than the PD.

\subsubsection{Effect of data generation conditions on performance}
In the data-generating models, the missingness mechanisms, proportions, skewness of the data, and correlation in the data were varied, resulting in 24 different simulation conditions. The plots displayed in Figure \ref{fig2}, \ref{fig3} and \ref{fig4} illustrate the impact of each of these conditions on the performance of the distance measures in terms of coverage, bias, and proportion of explained variance. The MCAR conditions show higher performance when compared to the MAR right conditions, and a higher proportion of missingness in the data leads to lower performance, as would be expected. The skewness of the data does not always impact the performance negatively. Under the MCAR conditions, a skewed distribution of the data results in higher coverage rates for some cases when compared to a normal distribution. Under the MAR right conditions, however, the opposite is true. As MAR right creates more missingness in the right tail of the data, and the data are positively skewed, the distribution of the data that are left disproportionately shifts to the left. This causes lower coverage rates and thus less valid inference. The proportion of explained variance (R squared) is relatively stable under the normal conditions but shows more variation under the skewed conditions. Finally, a higher correlation in the data under MCAR conditions does not lead to decreased performance, and in some cases to increased performance. It does lead to lower performance under the MAR right conditions. The proportion of explained variance increases with the correlation in the data, and is the largest under the MCAR condition with 25\% missingness, skewed distribution and a correlation of 0.7. 

\subsubsection{Effect of blending on performance}
The bias results in Figure \ref{fig2} show that under the conditions of MCAR, 50\% missingness and a normal distribution, the SBD with factor 0 and RBD with factors 0.5 and 0 slightly overestimate the predictions, whereas the PD and the blended metrics with higher blending factors slightly underestimate the predictions. In general, but especially in the more extreme simulation conditions, the predictions are underestimated by all metrics.

Figure \ref{fig3} shows that the trend for the coverage is similar across all conditions: the coverage rates become lower when the blended metrics are weighted more towards the MD. Under the conditions of MCAR, however, the results are similar for blending factors of 1 and 0.5, where a factor of 0.5 sometimes outperforms a factor of 1. The MD always leads to lower coverage rates. The downwards trend is especially pronounced under the conditions with a MAR right missingness mechanism, with coverage rates near only 40\% in the worst performing scenarios. 

Finally, the results for R squared in Figure \ref{fig4} show that the proportion of explained variance is relatively stable across the different metrics. Performance in terms of the explained variance is more dependent on the correlation in the data than on the distance measure used, as would be expected. Overall, however, it decreases as the blending factor decreases. Again, this trend is more strongly pronounced in the more extreme conditions.

There are a few exceptions where a blended distance measure with a blending factor of 0.5 or 0 outperforms either the PD, the blended distance measure with a blending factor of 1, or both. This is mostly the case under the MCAR conditions. It is important to note, however, that these differences are small and likely due to chance. The overall trend is that blending towards the MD leads to worse performance in bias, coverage and predictive power.

\section{Simulation study II}

\subsection{Aims}
The objective of the second simulation study is to provide a more detailed explanation of the results found in Simulation study I. In order to do so, I evaluate the version of the blended metric that performed best overall in Simulation study I, which is the ranked version. I simulate a setting where a single value is imputed, and evaluate the blended metric with blending factors ranging from 0 to 1, with intervals of 0.1.

\subsection{Data-generating mechanisms}
I expect that the distinction between the performance of the different blending factors will become more apparent under a more extreme data-generating mechanism. Therefore, the data are simulated from a single data-generating mechanism with the conditions under which the blended metric performed the worst in Simulation study I. The same variables are defined as in the first simulation study, where the data is skewed and the off-diagonal elements in the covariance are set to 0.7. A sample of size 500 is drawn. To simulate the prediction of the height measurement for a single target child, a random case in the data is made incomplete for the outcome.

\subsection{Estimand and performance measures}
The number of simulations is set to 10000. In every simulation, the outcome for a single random case is made incomplete and thereafter imputed 50 times. To assess the statistical properties of the PD and the RBD with different blending factors, I evaluate their accuracy, validity, and precision. For each simulation iteration, accuracy is evaluated by means of the (absolute) bias. This is done by evaluating the mean of the 50 imputations (the estimate) against the true value. In addition, accuracy is evaluated by the root mean square error (RMSE). Validity is evaluated by the coverage rate. Precision is evaluated in terms of the variance of the 50 imputations. 

\subsection{Results}
\begin{figure}[t]
\centerline{\includegraphics[width=\textwidth,height=\textheight,keepaspectratio]{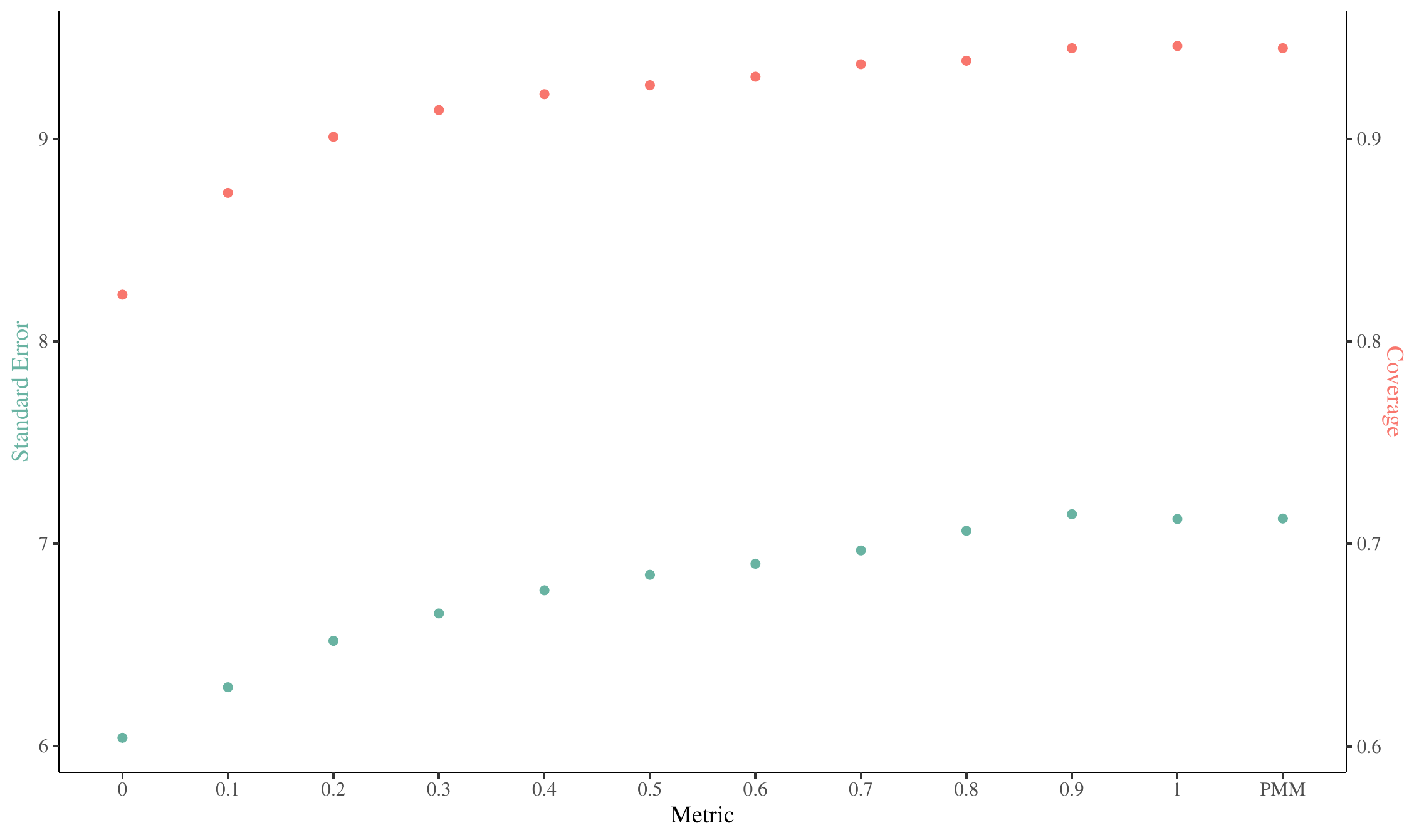}}
\caption{Average SE (green) and coverage (orange) simulation results for Simulation study II. The values on the x-axis indicate the blending factor used for the RBD. The left vertical axis indicates the values for the SE, the right vertical axis indicates the values for the coverage.\label{fig5}}
\end{figure}

\begin{figure}[t]
\centerline{\includegraphics[width=\textwidth,height=\textheight,keepaspectratio]{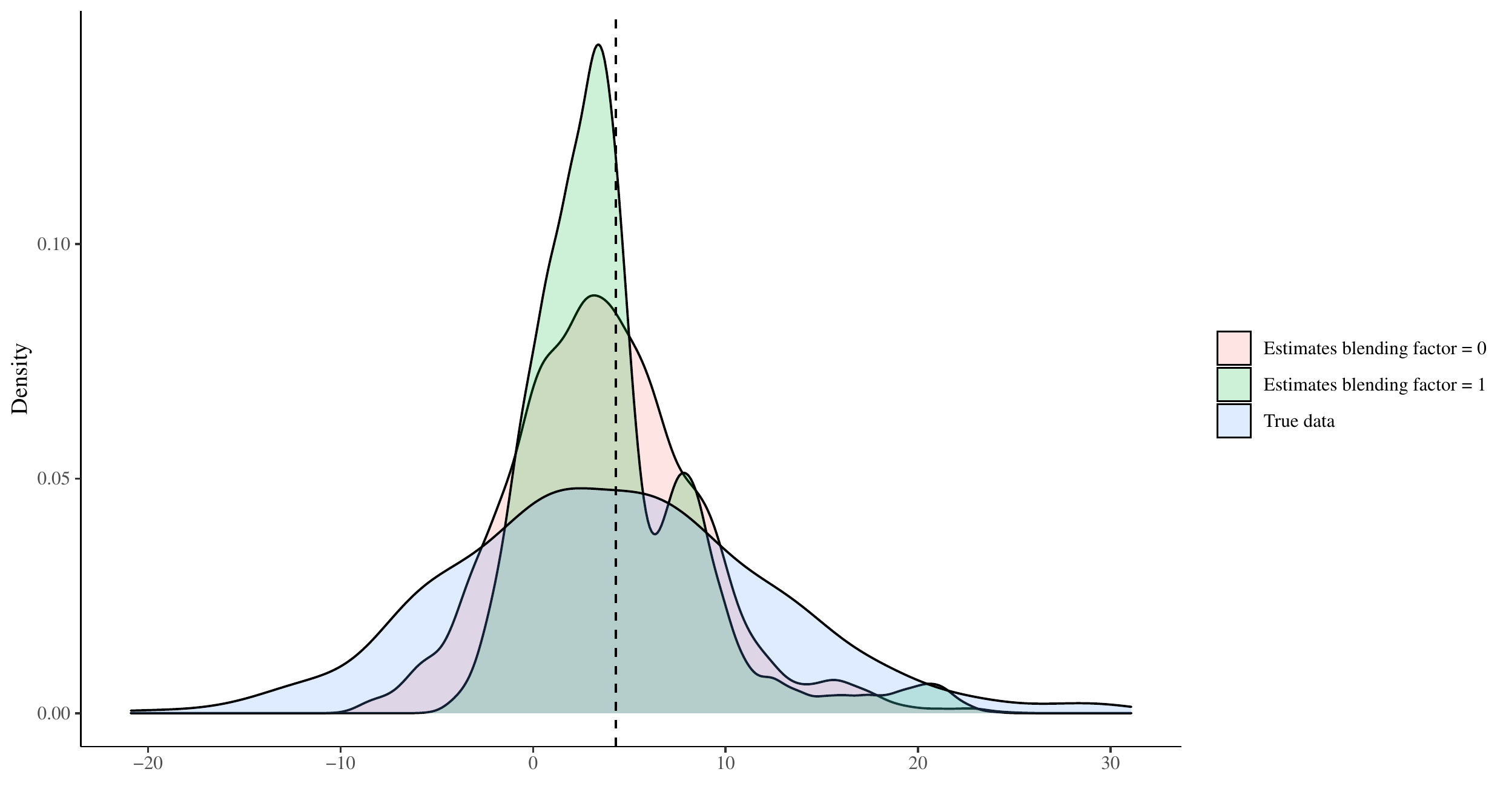}}
\caption{Density of the estimates obtained with the RBD with blending factor = 0 (MD) and of those obtained with the RBD with blending factor = 1 (PD), plotted against the density of the true data, e.g. the values that are made missing. The true estimate (expected value) is marked by the dashed line.\label{fig6}}
\end{figure}

\begin{figure}[t]
\centerline{\includegraphics[width=\textwidth,height=\textheight,keepaspectratio]{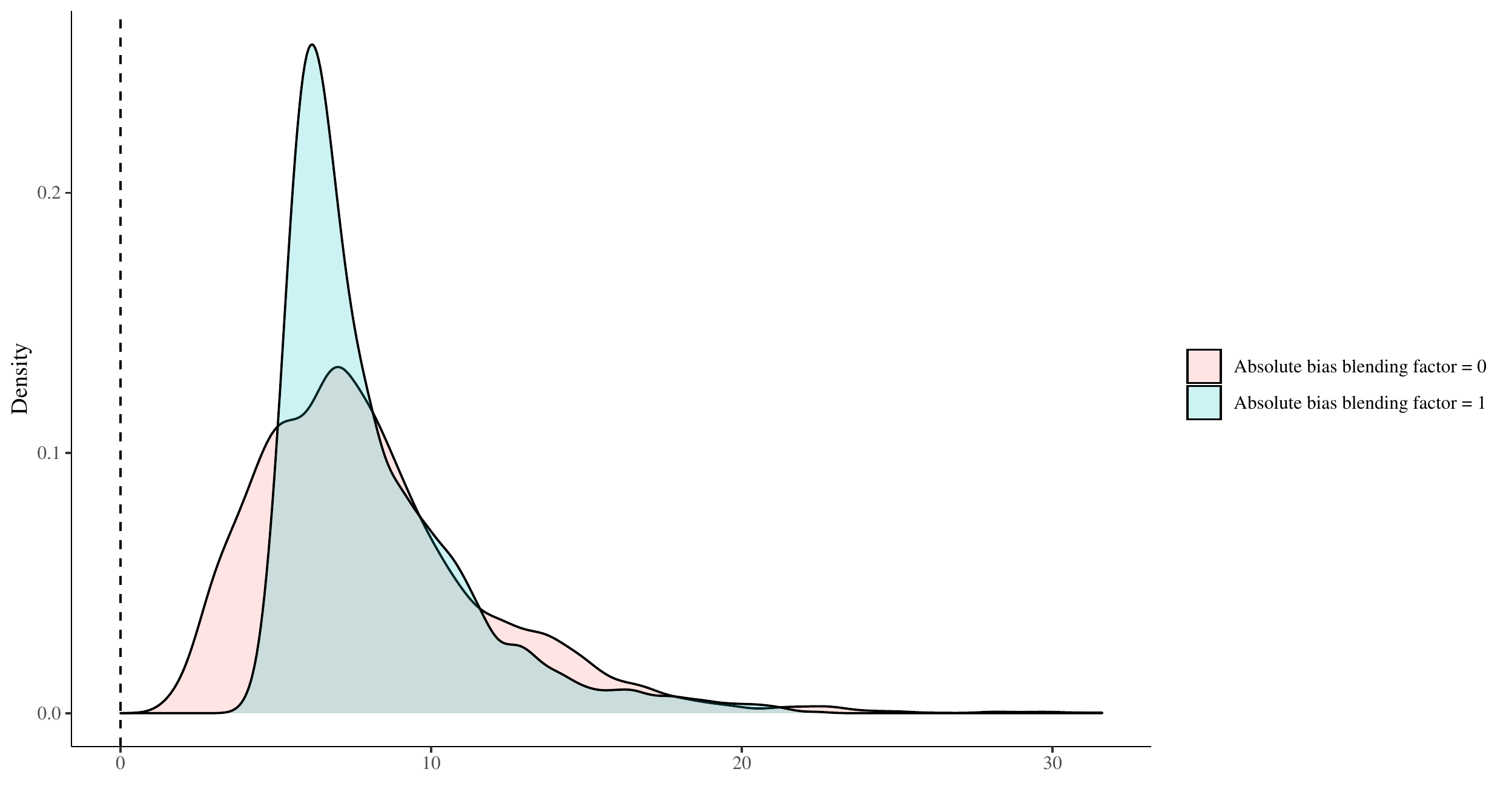}}
\caption{Density of the absolute bias results for the RBD with blending factor = 0 (MD) and of those obtained with the RBD with blending factor = 1 (PD). The dashed line marks an absolute bias of zero.\label{fig7}}
\end{figure}

Table \ref{tab8} displays the average of the performance results over the 10000 simulations. The results show that as the blending factor decreases (which implies weighting in the direction of the MD), the bias increases. The RMSE also shows a decreasing trend, however, there is a small increase in the RMSE from a blending factor of 0.1 to 0. There is no clear trend in the absolute bias. Figure \ref{fig5} shows that a higher blending factor leads to a larger standard error (SE) and a higher coverage rate. 
The performance measures show that a property of blending is the bias-variance trade off. That is, weighting towards the MD results in more precise, but less accurate estimates. The PD results in less precise, but more accurate estimates. Figure \ref{fig6} further illustrates this, which shows the density of the estimates obtained by the RBD with blending factor of 0 and a blending factor of 1, plotted against the density of the true data. A reference line is given at the expected true value, which is the point of evaluation. The PD has a higher density surrounding this point, indicating more certainty around the estimate. Figure \ref{fig7} shows the density of the absolute bias results for the same blending factors. Again, the high density in the plot shows that the PD causes estimates to be more certain.

\begin{table}[ht]
\centering
\caption{Performance results for Simulation study II. The table shows the average of the simulations for the estimate (mean of the 50 imputations), the true value of the missing outcome, (absolute) bias, sum of squared deviations (SSD), standard error (SE), lower and upper confidence limits, coverage, and root mean square error (RMSE).\label{tab8}}%
\begin{tabular}{lrrrrrrrrrr}
  \hline
method & estimate & true & bias & absbias & ssd & se & lwr & upr & cov & rmse \\ 
  \hline
PMM & 4.10 & 4.29 & -0.18 & 8.13 & 5133.00 & 7.12 & -10.21 & 18.42 & 0.94 & 9.68 \\ 
  Blending factor = 1 & 4.10 & 4.29 & -0.19 & 8.12 & 5126.81 & 7.12 & -10.22 & 18.41 & 0.95 & 9.67 \\ 
  Blending factor = 0.9 & 3.98 & 4.29 & -0.31 & 8.15 & 5182.68 & 7.15 & -10.38 & 18.34 & 0.94 & 9.70 \\ 
  Blending factor = 0.8 & 3.95 & 4.29 & -0.33 & 8.13 & 5153.59 & 7.06 & -10.24 & 18.15 & 0.94 & 9.65 \\ 
  Blending factor = 0.7 & 3.99 & 4.29 & -0.30 & 8.10 & 5116.94 & 6.97 & -10.01 & 17.99 & 0.94 & 9.60 \\ 
  Blending factor = 0.6 & 4.00 & 4.29 & -0.29 & 8.08 & 5107.43 & 6.90 & -9.87 & 17.87 & 0.93 & 9.57 \\ 
  Blending factor = 0.5 & 3.99 & 4.29 & -0.30 & 8.10 & 5118.92 & 6.85 & -9.77 & 17.75 & 0.93 & 9.57 \\ 
  Blending factor = 0.4 & 3.96 & 4.29 & -0.32 & 8.05 & 5082.51 & 6.77 & -9.64 & 17.57 & 0.92 & 9.51 \\ 
  Blending factor = 0.3 & 3.94 & 4.29 & -0.35 & 8.00 & 5025.93 & 6.66 & -9.43 & 17.32 & 0.91 & 9.43 \\ 
  Blending factor = 0.2 & 3.86 & 4.29 & -0.42 & 7.99 & 5014.08 & 6.52 & -9.24 & 16.97 & 0.90 & 9.38 \\ 
  Blending factor = 0.1 & 3.74 & 4.29 & -0.55 & 7.93 & 4948.25 & 6.29 & -8.91 & 16.38 & 0.87 & 9.25 \\ 
  Blending factor = 0 & 3.74 & 4.29 & -0.54 & 8.09 & 5121.67 & 6.04 & -8.40 & 15.89 & 0.82 & 9.32 \\ 
   \hline
\end{tabular}
\end{table}

\section{Discussion}
This study investigated the properties of a blended distance measure through two simulation studies. The purpose of Simulation study I was to evaluate the performance of the blended metric under different data-generating conditions. This simulation demonstrated that the blended metric performs worse when weighted more towards the MD, especially under extreme conditions like MAR right. The purpose of Simulation study II was to provide a further explanation of this result. The ranked version of the blended metric was evaluated under the condition of a skewed distribution and correlation of 0.7. The results show that a property of blending is the bias-variance trade off: weighting towards the PD results in less precise, but more accurate estimates. Furthermore, the coverage rates drop as the blended metric is weighted more towards the MD, which entails a decrease in statistical validity. 

It is more likely to select donors who are not people-like-me when using the PD. This can be counteracted by using the MD as a similarity measure, which causes there to be less variance among the selected donors, and thus leads to selection of people-like-me. However, this results in lower coverage rates, and therefore in less valid estimates. In practice, there is no need for concern about whether or not we select people-like-me, as the uncertainty is necessary for making valid inferences. 

The overall conclusion is that the blended metric can be implemented in situations where the missingness proportion is small and is MCAR. This is generally the case in the context of height prediction, as only a single value is to be imputed. It may be attractive for users of growth curve matching to use the blended metric when they have difficulty selecting a particular future time point to base matches on, or when they have more interest in the similarity between donors and the target in the predictor space. However, it is important to keep in mind the severe underperformance of the blended metric under more extreme conditions, and the reliable performance of the PD under any conditions. PMM should therefore always be the preferred method, and the results of this study may minimize doubts about discrepancies between the trajectories of the matched donors and the target. After all, our past does not always define our future, and selecting appropriate matches may not so much be a matter of finding people-like-me in terms of historic growth, as it is one of future similarity. 

The current study investigated the influence of missingness proportion, missingness mechanisms, skewness of the data, and correlation in the data for the use of the blended metric. A limitation is that the blended metric might perform better in practice, and therefore it would be useful to evaluate the metric in empirical data as well. Additionally, alternative combinations of similarity measures and the PD could lead to different results, and these combinations were not studied. Examples of other similarity measures would be the Fréchet distance, \cite{eiter_computing_1994} and the locally supervised metric learning (LSML) measure.\cite{ng_personalized_2015} Finally, further research could investigate the impact of more variations in simulation conditions such as the sample size and the number of \textit{k} matched donors, and could determine what the optimal blending factor is to predict outcomes.

\clearpage

\section*{Supporting information}
Instructions and scripts to reproduce the simulation results are available in the \href{https://github.com/anaisfopma/Blended-Distance-Thesis}{research archive} of this project. The study was approved by the Ethical Review Board of the Faculty of Social and Behavioural Sciences of Utrecht University. The approval is based on the documents sent by the researchers as requested in the form of the Ethics committee and filed under number 21-1906.

\section*{Acknowledgments}
I would like to thank Gerko Vink, Mingyang Cai, and Stef van Buuren for their guidance and their supervision of the thesis. Furthermore, I would like to thank the other members of the missing data team, in particular Hanne Oberman and Thom Volker, for their helpful comments and feedback. I appreciate that I could join the MICE meetings and have learned a lot from the talented people that make up this team. Lastly, I would like to thank all my peers in the MSBBSS 2022 cohort, both for their feedback during the research seminars and their support during this last semester of the master’s programme.

\bibliographystyle{unsrt}  
\bibliography{references}

\begin{thebibliography}{10}

\bibitem{straatmann_how_2018}
Viviane~S. Straatmann, Anna Pearce, Steven Hope, Benjamin Barr, Margaret
  Whitehead, Catherine Law, and David Taylor-Robinson.
\newblock How well can poor child health and development be predicted by data
  collected in early childhood?
\newblock {\em J Epidemiol Community Health}, 72(12):1132--1140, 2018.
\newblock Publisher: BMJ Publishing Group Ltd.

\bibitem{cordeiro_childs_2019}
João~Rala Cordeiro, Octavian Postolache, and João~C. Ferreira.
\newblock Child’s target height prediction evolution.
\newblock {\em Applied Sciences}, 9(24):5447, 2019.
\newblock Publisher: Multidisciplinary Digital Publishing Institute.

\bibitem{van_buuren_curve_2014}
Stef Van~Buuren.
\newblock Curve matching: a data-driven technique to improve individual
  prediction of childhood growth.
\newblock {\em Annals of Nutrition and Metabolism}, 65(2-3):227--233, 2014.
\newblock Publisher: Karger Publishers.

\bibitem{van_buuren_broken_2020}
Stef Van~Buuren.
\newblock Broken stick model for irregular longitudinal data.
\newblock {\em Journal of Statistical Software}, Submitted for
  publication:1--47, 2020.

\bibitem{herngreen_growth_1994}
W.~P. Herngreen, Stef Van~Buuren, J.~C. Van~Wieringen, J.~D. Reerink, S.~P.
  Verloove-Vanhorick, and J.~H. Ruys.
\newblock Growth in length and weight from birth to 2 years of a representative
  sample of {Netherlands} children (born in 1988–89) related to socioeconomic
  status and other background characteristics.
\newblock {\em Annals of human biology}, 21(5):449--463, 1994.
\newblock Publisher: Taylor \& Francis.

\bibitem{morris_using_2019}
Tim~P. Morris, Ian~R. White, and Michael~J. Crowther.
\newblock Using simulation studies to evaluate statistical methods.
\newblock {\em Statistics in Medicine}, 38(11):2074--2102, 2019.
\newblock \_eprint: https://onlinelibrary.wiley.com/doi/pdf/10.1002/sim.8086.

\bibitem{R2022}
{R Core Team}.
\newblock {\em R: A Language and Environment for Statistical Computing}.
\newblock R Foundation for Statistical Computing, Vienna, Austria, 2022.

\bibitem{van_buuren_mice_2011}
Stef {van Buuren} and Karin Groothuis-Oudshoorn.
\newblock {mice}: Multivariate imputation by chained equations in r.
\newblock {\em Journal of Statistical Software}, 45(3):1--67, 2011.

\bibitem{vink_predictive_2014}
Gerko Vink, Laurence~E. Frank, Jeroen Pannekoek, and Stef Van~Buuren.
\newblock Predictive mean matching imputation of semicontinuous variables.
\newblock {\em Statistica Neerlandica}, 68(1):61--90, 2014.
\newblock Publisher: Wiley Online Library.

\bibitem{little_statistical_2019}
Roderick~JA Little and Donald~B. Rubin.
\newblock {\em Statistical analysis with missing data}, volume 793.
\newblock John Wiley \& Sons, 2019.

\bibitem{van_buuren_flexible_2018}
Stef Van~Buuren.
\newblock {\em Flexible imputation of missing data}.
\newblock CRC press, 2018.

\bibitem{vink_pooling_2014}
Gerko Vink and Stef van Buuren.
\newblock Pooling multiple imputations when the sample happens to be the
  population.
\newblock {\em arXiv preprint arXiv:1409.8542}, 2014.

\bibitem{eiter_computing_1994}
Thomas Eiter and Heikki Mannila.
\newblock Computing discrete {Fréchet} distance.
\newblock Technical report, Citeseer, 1994.

\bibitem{ng_personalized_2015}
Kenney Ng, Jimeng Sun, Jianying Hu, and Fei Wang.
\newblock Personalized predictive modeling and risk factor identification using
  patient similarity.
\newblock {\em AMIA Summits on Translational Science Proceedings}, 2015:132,
  2015.
\newblock Publisher: American Medical Informatics Association.

\end{thebibliography}

\clearpage
\appendix

\section{Results of Simulation study I}\label{app1}

\begin{sidewaystable}
\centering
\caption{Method PMM.\label{tab1}}%
\begin{tabular}{lllllllllllllll}
  \hline
mech & mis & dist & cor & qbar & se & t & df & b & 2.5\% & 97.5\% & true & cov & bias & R2 \\ 
  \hline
MCAR & 25\% & normal & 0 & 30.466 & 0.191 & 0.153 & 97.773 & 0.035 & 29.935 & 30.996 & 30.47 & 0.949 & -0.004 & 0.079 \\ 
   &  &  & 0.1 & 30.466 & 0.192 & 0.155 & 98.311 & 0.035 & 29.933 & 30.998 & 30.475 & 0.96 & -0.009 & 0.09 \\ 
   &  &  & 0.7 & 30.502 & 0.187 & 0.161 & 112.474 & 0.033 & 29.982 & 31.022 & 30.502 & 0.941 & 0 & 0.152 \\ 
   &  & skewed & 0 & 3.199 & 0.248 & 0.187 & 65.411 & 0.06 & 2.511 & 3.888 & 3.256 & 0.969 & -0.057 & 0.106 \\ 
   &  &  & 0.1 & 3.238 & 0.251 & 0.189 & 64.18 & 0.061 & 2.543 & 3.934 & 3.292 & 0.976 & -0.054 & 0.112 \\ 
   &  &  & 0.7 & 4.068 & 0.259 & 0.213 & 67.895 & 0.064 & 3.349 & 4.788 & 4.109 & 0.976 & -0.041 & 0.239 \\ 
   & 50\% & normal & 0 & 30.443 & 0.339 & 0.243 & 30.409 & 0.109 & 29.502 & 31.383 & 30.47 & 0.94 & -0.027 & 0.083 \\ 
   &  &  & 0.1 & 30.45 & 0.342 & 0.248 & 31.72 & 0.113 & 29.499 & 31.4 & 30.475 & 0.953 & -0.026 & 0.093 \\ 
   &  &  & 0.7 & 30.495 & 0.327 & 0.243 & 35.429 & 0.102 & 29.587 & 31.403 & 30.502 & 0.94 & -0.007 & 0.153 \\ 
   &  & skewed & 0 & 3.149 & 0.462 & 0.373 & 21.577 & 0.214 & 1.867 & 4.431 & 3.256 & 0.977 & -0.107 & 0.106 \\ 
   &  &  & 0.1 & 3.167 & 0.49 & 0.402 & 17.865 & 0.238 & 1.807 & 4.526 & 3.292 & 0.978 & -0.126 & 0.114 \\ 
   &  &  & 0.7 & 4.022 & 0.516 & 0.444 & 18.404 & 0.257 & 2.59 & 5.455 & 4.109 & 0.978 & -0.087 & 0.24 \\ 
  MAR & 25\% & normal & 0 & 30.456 & 0.185 & 0.151 & 105.503 & 0.033 & 29.943 & 30.969 & 30.47 & 0.912 & -0.014 & 0.082 \\ 
   &  &  & 0.1 & 30.46 & 0.186 & 0.153 & 105.455 & 0.033 & 29.943 & 30.977 & 30.475 & 0.897 & -0.015 & 0.093 \\ 
   &  &  & 0.7 & 30.483 & 0.188 & 0.162 & 113.281 & 0.034 & 29.96 & 31.005 & 30.502 & 0.908 & -0.019 & 0.153 \\ 
   &  & skewed & 0 & 3.09 & 0.193 & 0.154 & 101.507 & 0.036 & 2.556 & 3.625 & 3.256 & 0.878 & -0.165 & 0.077 \\ 
   &  &  & 0.1 & 3.143 & 0.195 & 0.156 & 96.792 & 0.036 & 2.601 & 3.685 & 3.292 & 0.884 & -0.149 & 0.088 \\ 
   &  &  & 0.7 & 3.895 & 0.194 & 0.17 & 107.547 & 0.036 & 3.356 & 4.434 & 4.109 & 0.842 & -0.214 & 0.176 \\ 
   & 50\% & normal & 0 & 30.437 & 0.303 & 0.218 & 41.511 & 0.089 & 29.596 & 31.277 & 30.47 & 0.863 & -0.033 & 0.086 \\ 
   &  &  & 0.1 & 30.446 & 0.305 & 0.221 & 37.774 & 0.09 & 29.599 & 31.292 & 30.475 & 0.868 & -0.03 & 0.097 \\ 
   &  &  & 0.7 & 30.451 & 0.305 & 0.228 & 43.159 & 0.09 & 29.605 & 31.296 & 30.502 & 0.857 & -0.051 & 0.153 \\ 
   &  & skewed & 0 & 2.975 & 0.372 & 0.271 & 26.075 & 0.133 & 1.942 & 4.008 & 3.256 & 0.911 & -0.281 & 0.081 \\ 
   &  &  & 0.1 & 3.021 & 0.365 & 0.267 & 26.939 & 0.129 & 2.007 & 4.035 & 3.292 & 0.905 & -0.272 & 0.089 \\ 
   &  &  & 0.7 & 3.674 & 0.37 & 0.28 & 28.378 & 0.132 & 2.646 & 4.702 & 4.109 & 0.858 & -0.436 & 0.141 \\ 
   \hline
\end{tabular}
\end{sidewaystable}

\begin{sidewaystable}
\centering
\caption{Method ranked, blend = 1.\label{tab2}}%
\begin{tabular}{lllllllllllllll}
  \hline
mech & mis & dist & cor & qbar & se & t & df & b & 2.5\% & 97.5\% & true & cov & bias & R2 \\ 
  \hline
MCAR & 25\% & normal & 0 & 30.462 & 0.192 & 0.154 & 97.885 & 0.035 & 29.929 & 30.995 & 30.47 & 0.95 & -0.008 & 0.079 \\ 
   &  &  & 0.1 & 30.467 & 0.195 & 0.156 & 95.776 & 0.036 & 29.927 & 31.008 & 30.475 & 0.961 & -0.008 & 0.091 \\ 
   &  &  & 0.7 & 30.504 & 0.185 & 0.16 & 111.671 & 0.033 & 29.989 & 31.019 & 30.502 & 0.952 & 0.003 & 0.152 \\ 
   &  & skewed & 0 & 3.211 & 0.242 & 0.183 & 66.725 & 0.056 & 2.539 & 3.883 & 3.256 & 0.975 & -0.045 & 0.107 \\ 
   &  &  & 0.1 & 3.234 & 0.254 & 0.191 & 60.949 & 0.062 & 2.529 & 3.939 & 3.292 & 0.975 & -0.058 & 0.112 \\ 
   &  &  & 0.7 & 4.074 & 0.253 & 0.21 & 71.264 & 0.062 & 3.37 & 4.778 & 4.109 & 0.968 & -0.035 & 0.24 \\ 
   & 50\% & normal & 0 & 30.44 & 0.34 & 0.244 & 30.283 & 0.11 & 29.496 & 31.384 & 30.47 & 0.949 & -0.03 & 0.083 \\ 
   &  &  & 0.1 & 30.454 & 0.329 & 0.238 & 35.708 & 0.104 & 29.541 & 31.366 & 30.475 & 0.93 & -0.022 & 0.094 \\ 
   &  &  & 0.7 & 30.492 & 0.328 & 0.244 & 35.911 & 0.103 & 29.58 & 31.404 & 30.502 & 0.939 & -0.009 & 0.153 \\ 
   &  & skewed & 0 & 3.147 & 0.461 & 0.375 & 21.325 & 0.217 & 1.866 & 4.428 & 3.256 & 0.965 & -0.109 & 0.106 \\ 
   &  &  & 0.1 & 3.155 & 0.487 & 0.403 & 19.639 & 0.238 & 1.803 & 4.507 & 3.292 & 0.972 & -0.137 & 0.115 \\ 
   &  &  & 0.7 & 4.014 & 0.513 & 0.442 & 17.427 & 0.256 & 2.589 & 5.438 & 4.109 & 0.982 & -0.096 & 0.24 \\ 
  MAR & 25\% & normal & 0 & 30.46 & 0.185 & 0.151 & 105.089 & 0.032 & 29.946 & 30.973 & 30.47 & 0.898 & -0.01 & 0.082 \\ 
   &  &  & 0.1 & 30.458 & 0.188 & 0.153 & 103.006 & 0.033 & 29.937 & 30.979 & 30.475 & 0.898 & -0.017 & 0.094 \\ 
   &  &  & 0.7 & 30.481 & 0.188 & 0.162 & 110.033 & 0.034 & 29.958 & 31.004 & 30.502 & 0.902 & -0.021 & 0.153 \\ 
   &  & skewed & 0 & 3.088 & 0.191 & 0.153 & 100.427 & 0.035 & 2.558 & 3.619 & 3.256 & 0.866 & -0.167 & 0.077 \\ 
   &  &  & 0.1 & 3.145 & 0.193 & 0.155 & 98.57 & 0.035 & 2.611 & 3.68 & 3.292 & 0.884 & -0.147 & 0.088 \\ 
   &  &  & 0.7 & 3.892 & 0.192 & 0.169 & 107.928 & 0.035 & 3.358 & 4.426 & 4.109 & 0.825 & -0.218 & 0.175 \\ 
   & 50\% & normal & 0 & 30.444 & 0.301 & 0.218 & 41.213 & 0.088 & 29.607 & 31.281 & 30.47 & 0.852 & -0.025 & 0.086 \\ 
   &  &  & 0.1 & 30.442 & 0.304 & 0.22 & 41.027 & 0.089 & 29.598 & 31.286 & 30.475 & 0.883 & -0.033 & 0.097 \\ 
   &  &  & 0.7 & 30.459 & 0.301 & 0.226 & 45.189 & 0.087 & 29.623 & 31.295 & 30.502 & 0.865 & -0.043 & 0.153 \\ 
   &  & skewed & 0 & 2.983 & 0.381 & 0.279 & 24.606 & 0.14 & 1.923 & 4.042 & 3.256 & 0.918 & -0.273 & 0.08 \\ 
   &  &  & 0.1 & 3.025 & 0.372 & 0.273 & 27.808 & 0.134 & 1.994 & 4.057 & 3.292 & 0.901 & -0.267 & 0.089 \\ 
   &  &  & 0.7 & 3.665 & 0.367 & 0.276 & 27.827 & 0.129 & 2.646 & 4.684 & 4.109 & 0.836 & -0.444 & 0.142 \\ 
   \hline
\end{tabular}
\end{sidewaystable}

\begin{sidewaystable}
\centering
\caption{Method ranked, blend = 0.5.\label{tab3}}%
\begin{tabular}{lllllllllllllll}
  \hline
mech & mis & dist & cor & qbar & se & t & df & b & 2.5\% & 97.5\% & true & cov & bias & R2 \\ 
  \hline
MCAR & 25\% & normal & 0 & 30.477 & 0.181 & 0.149 & 107.792 & 0.031 & 29.974 & 30.98 & 30.47 & 0.957 & 0.007 & 0.076 \\ 
   &  &  & 0.1 & 30.492 & 0.183 & 0.151 & 106.919 & 0.031 & 29.985 & 30.999 & 30.475 & 0.943 & 0.016 & 0.088 \\ 
   &  &  & 0.7 & 30.51 & 0.18 & 0.158 & 119.4 & 0.031 & 30.01 & 31.009 & 30.502 & 0.938 & 0.008 & 0.148 \\ 
   &  & skewed & 0 & 3.227 & 0.204 & 0.161 & 90.943 & 0.04 & 2.662 & 3.792 & 3.256 & 0.944 & -0.029 & 0.094 \\ 
   &  &  & 0.1 & 3.264 & 0.214 & 0.167 & 84.099 & 0.044 & 2.67 & 3.857 & 3.292 & 0.951 & -0.029 & 0.102 \\ 
   &  &  & 0.7 & 4.05 & 0.224 & 0.191 & 85.852 & 0.048 & 3.426 & 4.673 & 4.109 & 0.95 & -0.06 & 0.226 \\ 
   & 50\% & normal & 0 & 30.486 & 0.303 & 0.216 & 38.173 & 0.087 & 29.646 & 31.326 & 30.47 & 0.917 & 0.017 & 0.074 \\ 
   &  &  & 0.1 & 30.494 & 0.296 & 0.212 & 39.051 & 0.083 & 29.673 & 31.315 & 30.475 & 0.928 & 0.019 & 0.085 \\ 
   &  &  & 0.7 & 30.516 & 0.293 & 0.219 & 47.093 & 0.082 & 29.704 & 31.328 & 30.502 & 0.914 & 0.015 & 0.142 \\ 
   &  & skewed & 0 & 3.178 & 0.351 & 0.256 & 31.295 & 0.12 & 2.202 & 4.154 & 3.256 & 0.946 & -0.078 & 0.087 \\ 
   &  &  & 0.1 & 3.218 & 0.37 & 0.271 & 26.642 & 0.131 & 2.191 & 4.245 & 3.292 & 0.946 & -0.074 & 0.094 \\ 
   &  &  & 0.7 & 3.941 & 0.409 & 0.325 & 26.368 & 0.161 & 2.805 & 5.076 & 4.109 & 0.951 & -0.169 & 0.206 \\ 
  MAR & 25\% & normal & 0 & 30.421 & 0.178 & 0.148 & 111.804 & 0.03 & 29.927 & 30.915 & 30.47 & 0.898 & -0.049 & 0.076 \\ 
   &  &  & 0.1 & 30.421 & 0.177 & 0.149 & 116.968 & 0.03 & 29.93 & 30.912 & 30.475 & 0.898 & -0.054 & 0.088 \\ 
   &  &  & 0.7 & 30.434 & 0.18 & 0.158 & 117.612 & 0.031 & 29.934 & 30.934 & 30.502 & 0.898 & -0.068 & 0.145 \\ 
   &  & skewed & 0 & 3.081 & 0.177 & 0.146 & 114.292 & 0.03 & 2.589 & 3.573 & 3.256 & 0.864 & -0.175 & 0.07 \\ 
   &  &  & 0.1 & 3.109 & 0.175 & 0.146 & 118.274 & 0.029 & 2.624 & 3.594 & 3.292 & 0.847 & -0.183 & 0.078 \\ 
   &  &  & 0.7 & 3.76 & 0.17 & 0.155 & 133.762 & 0.028 & 3.287 & 4.233 & 4.109 & 0.678 & -0.35 & 0.141 \\ 
   & 50\% & normal & 0 & 30.406 & 0.262 & 0.191 & 52.803 & 0.066 & 29.678 & 31.133 & 30.47 & 0.809 & -0.064 & 0.082 \\ 
   &  &  & 0.1 & 30.402 & 0.273 & 0.199 & 48.497 & 0.071 & 29.643 & 31.161 & 30.475 & 0.84 & -0.073 & 0.092 \\ 
   &  &  & 0.7 & 30.396 & 0.264 & 0.202 & 58.445 & 0.068 & 29.662 & 31.13 & 30.502 & 0.802 & -0.105 & 0.145 \\ 
   &  & skewed & 0 & 2.907 & 0.301 & 0.213 & 37.629 & 0.087 & 2.071 & 3.744 & 3.256 & 0.822 & -0.348 & 0.059 \\ 
   &  &  & 0.1 & 2.932 & 0.293 & 0.207 & 41.762 & 0.081 & 2.12 & 3.744 & 3.292 & 0.801 & -0.361 & 0.065 \\ 
   &  &  & 0.7 & 3.547 & 0.29 & 0.214 & 46.431 & 0.081 & 2.741 & 4.353 & 4.109 & 0.687 & -0.563 & 0.109 \\ 
   \hline
\end{tabular}
\end{sidewaystable}

\begin{sidewaystable}
\centering
\caption{Method ranked, blend = 0.\label{tab4}}%
\begin{tabular}{lllllllllllllll}
  \hline
mech & mis & dist & cor & qbar & se & t & df & b & 2.5\% & 97.5\% & true & cov & bias & R2 \\ 
  \hline
MCAR & 25\% & normal & 0 & 30.462 & 0.145 & 0.134 & 156.897 & 0.02 & 30.059 & 30.866 & 30.47 & 0.888 & -0.007 & 0.074 \\ 
   &  &  & 0.1 & 30.468 & 0.146 & 0.136 & 159.425 & 0.02 & 30.062 & 30.874 & 30.475 & 0.893 & -0.007 & 0.086 \\ 
   &  &  & 0.7 & 30.499 & 0.148 & 0.145 & 166.98 & 0.021 & 30.088 & 30.91 & 30.502 & 0.882 & -0.003 & 0.144 \\ 
   &  & skewed & 0 & 3.192 & 0.147 & 0.138 & 158.72 & 0.021 & 2.782 & 3.601 & 3.256 & 0.893 & -0.064 & 0.097 \\ 
   &  &  & 0.1 & 3.223 & 0.147 & 0.139 & 162.631 & 0.02 & 2.816 & 3.63 & 3.292 & 0.889 & -0.069 & 0.103 \\ 
   &  &  & 0.7 & 4.003 & 0.146 & 0.157 & 186.987 & 0.02 & 3.599 & 4.407 & 4.109 & 0.863 & -0.106 & 0.223 \\ 
   & 50\% & normal & 0 & 30.494 & 0.211 & 0.159 & 77.005 & 0.042 & 29.909 & 31.079 & 30.47 & 0.839 & 0.024 & 0.069 \\ 
   &  &  & 0.1 & 30.504 & 0.211 & 0.161 & 77.508 & 0.042 & 29.918 & 31.091 & 30.475 & 0.851 & 0.029 & 0.079 \\ 
   &  &  & 0.7 & 30.531 & 0.212 & 0.169 & 87.214 & 0.043 & 29.944 & 31.118 & 30.502 & 0.843 & 0.029 & 0.132 \\ 
   &  & skewed & 0 & 3.128 & 0.209 & 0.16 & 79.65 & 0.041 & 2.548 & 3.707 & 3.256 & 0.826 & -0.128 & 0.091 \\ 
   &  &  & 0.1 & 3.146 & 0.207 & 0.161 & 82.816 & 0.04 & 2.571 & 3.721 & 3.292 & 0.835 & -0.146 & 0.097 \\ 
   &  &  & 0.7 & 3.846 & 0.213 & 0.181 & 92.397 & 0.043 & 3.255 & 4.438 & 4.109 & 0.772 & -0.263 & 0.2 \\ 
  MAR & 25\% & normal & 0 & 30.408 & 0.15 & 0.136 & 148.803 & 0.021 & 29.99 & 30.825 & 30.47 & 0.863 & -0.062 & 0.078 \\ 
   &  &  & 0.1 & 30.401 & 0.151 & 0.138 & 152.097 & 0.021 & 29.983 & 30.819 & 30.475 & 0.87 & -0.074 & 0.088 \\ 
   &  &  & 0.7 & 30.402 & 0.149 & 0.145 & 160.168 & 0.021 & 29.987 & 30.817 & 30.502 & 0.84 & -0.1 & 0.144 \\ 
   &  & skewed & 0 & 3.054 & 0.145 & 0.134 & 158.032 & 0.02 & 2.651 & 3.457 & 3.256 & 0.753 & -0.202 & 0.07 \\ 
   &  &  & 0.1 & 3.088 & 0.147 & 0.135 & 156.613 & 0.02 & 2.681 & 3.496 & 3.292 & 0.758 & -0.204 & 0.074 \\ 
   &  &  & 0.7 & 3.681 & 0.143 & 0.143 & 174.906 & 0.019 & 3.284 & 4.077 & 4.109 & 0.453 & -0.429 & 0.13 \\ 
   & 50\% & normal & 0 & 30.357 & 0.214 & 0.163 & 77.74 & 0.044 & 29.762 & 30.952 & 30.47 & 0.761 & -0.113 & 0.08 \\ 
   &  &  & 0.1 & 30.349 & 0.215 & 0.164 & 74.64 & 0.044 & 29.751 & 30.947 & 30.475 & 0.761 & -0.126 & 0.089 \\ 
   &  &  & 0.7 & 30.318 & 0.214 & 0.171 & 82.206 & 0.043 & 29.724 & 30.911 & 30.502 & 0.736 & -0.184 & 0.141 \\ 
   &  & skewed & 0 & 2.86 & 0.205 & 0.154 & 78.597 & 0.04 & 2.29 & 3.43 & 3.256 & 0.651 & -0.396 & 0.059 \\ 
   &  &  & 0.1 & 2.895 & 0.201 & 0.154 & 84.367 & 0.038 & 2.336 & 3.455 & 3.292 & 0.638 & -0.397 & 0.065 \\ 
   &  &  & 0.7 & 3.44 & 0.208 & 0.165 & 84.385 & 0.041 & 2.863 & 4.018 & 4.109 & 0.403 & -0.669 & 0.1 \\ 
   \hline
\end{tabular}
\end{sidewaystable}

\begin{sidewaystable}
\centering
\caption{Method scaled, blend = 1.\label{tab5}}%
\begin{tabular}{lllllllllllllll}
  \hline
mech & mis & dist & cor & qbar & se & t & df & b & 2.5\% & 97.5\% & true & cov & bias & R2 \\ 
  \hline
MCAR & 25\% & normal & 0 & 30.461 & 0.197 & 0.156 & 93.518 & 0.037 & 29.914 & 31.009 & 30.47 & 0.958 & -0.008 & 0.079 \\ 
   &  &  & 0.1 & 30.473 & 0.189 & 0.154 & 103.355 & 0.034 & 29.947 & 30.999 & 30.475 & 0.942 & -0.002 & 0.09 \\ 
   &  &  & 0.7 & 30.507 & 0.194 & 0.164 & 100.127 & 0.035 & 29.968 & 31.046 & 30.502 & 0.964 & 0.005 & 0.152 \\ 
   &  & skewed & 0 & 3.205 & 0.25 & 0.188 & 60.828 & 0.06 & 2.51 & 3.901 & 3.256 & 0.968 & -0.05 & 0.107 \\ 
   &  &  & 0.1 & 3.236 & 0.248 & 0.188 & 62.727 & 0.059 & 2.548 & 3.924 & 3.292 & 0.969 & -0.056 & 0.112 \\ 
   &  &  & 0.7 & 4.073 & 0.255 & 0.211 & 72.87 & 0.063 & 3.365 & 4.782 & 4.109 & 0.966 & -0.036 & 0.24 \\ 
   & 50\% & normal & 0 & 30.439 & 0.338 & 0.243 & 31.954 & 0.11 & 29.501 & 31.376 & 30.47 & 0.939 & -0.031 & 0.084 \\ 
   &  &  & 0.1 & 30.45 & 0.33 & 0.24 & 35.211 & 0.106 & 29.534 & 31.367 & 30.475 & 0.936 & -0.025 & 0.093 \\ 
   &  &  & 0.7 & 30.488 & 0.33 & 0.245 & 35.929 & 0.103 & 29.572 & 31.405 & 30.502 & 0.948 & -0.013 & 0.153 \\ 
   &  & skewed & 0 & 3.152 & 0.458 & 0.371 & 22.021 & 0.213 & 1.882 & 4.423 & 3.256 & 0.969 & -0.103 & 0.107 \\ 
   &  &  & 0.1 & 3.171 & 0.484 & 0.398 & 18.633 & 0.234 & 1.826 & 4.515 & 3.292 & 0.978 & -0.122 & 0.113 \\ 
   &  &  & 0.7 & 4.013 & 0.518 & 0.447 & 17.036 & 0.259 & 2.575 & 5.451 & 4.109 & 0.987 & -0.096 & 0.241 \\ 
  MAR & 25\% & normal & 0 & 30.461 & 0.185 & 0.151 & 104.176 & 0.032 & 29.947 & 30.975 & 30.47 & 0.906 & -0.009 & 0.082 \\ 
   &  &  & 0.1 & 30.46 & 0.189 & 0.154 & 102.232 & 0.034 & 29.935 & 30.984 & 30.475 & 0.916 & -0.016 & 0.093 \\ 
   &  &  & 0.7 & 30.489 & 0.188 & 0.162 & 109.703 & 0.033 & 29.968 & 31.011 & 30.502 & 0.908 & -0.012 & 0.154 \\ 
   &  & skewed & 0 & 3.094 & 0.191 & 0.153 & 104.554 & 0.036 & 2.563 & 3.626 & 3.256 & 0.868 & -0.162 & 0.077 \\ 
   &  &  & 0.1 & 3.14 & 0.193 & 0.155 & 96.462 & 0.036 & 2.603 & 3.677 & 3.292 & 0.867 & -0.152 & 0.088 \\ 
   &  &  & 0.7 & 3.889 & 0.193 & 0.17 & 111.041 & 0.035 & 3.354 & 4.424 & 4.109 & 0.825 & -0.221 & 0.175 \\ 
   & 50\% & normal & 0 & 30.439 & 0.299 & 0.217 & 44.292 & 0.088 & 29.608 & 31.27 & 30.47 & 0.859 & -0.031 & 0.086 \\ 
   &  &  & 0.1 & 30.441 & 0.301 & 0.218 & 39.103 & 0.088 & 29.604 & 31.278 & 30.475 & 0.866 & -0.034 & 0.097 \\ 
   &  &  & 0.7 & 30.457 & 0.296 & 0.222 & 46.435 & 0.085 & 29.635 & 31.278 & 30.502 & 0.857 & -0.045 & 0.153 \\ 
   &  & skewed & 0 & 2.994 & 0.378 & 0.279 & 27.141 & 0.14 & 1.944 & 4.044 & 3.256 & 0.904 & -0.262 & 0.08 \\ 
   &  &  & 0.1 & 3.029 & 0.374 & 0.275 & 25.454 & 0.136 & 1.99 & 4.068 & 3.292 & 0.906 & -0.263 & 0.089 \\ 
   &  &  & 0.7 & 3.661 & 0.371 & 0.278 & 27.739 & 0.131 & 2.632 & 4.69 & 4.109 & 0.851 & -0.448 & 0.141 \\ 
   \hline
\end{tabular}
\end{sidewaystable}

\begin{sidewaystable}
\centering
\caption{Method scaled, blend = 0.5.\label{tab6}}%
\begin{tabular}{lllllllllllllll}
  \hline
mech & mis & dist & cor & qbar & se & t & df & b & 2.5\% & 97.5\% & true & cov & bias & R2 \\ 
  \hline
MCAR & 25\% & normal & 0 & 30.467 & 0.188 & 0.152 & 102.834 & 0.034 & 29.946 & 30.988 & 30.47 & 0.953 & -0.003 & 0.077 \\ 
   &  &  & 0.1 & 30.478 & 0.18 & 0.15 & 112.559 & 0.031 & 29.979 & 30.978 & 30.475 & 0.94 & 0.003 & 0.088 \\ 
   &  &  & 0.7 & 30.496 & 0.178 & 0.158 & 125.64 & 0.03 & 30.002 & 30.99 & 30.502 & 0.939 & -0.006 & 0.149 \\ 
   &  & skewed & 0 & 3.184 & 0.246 & 0.186 & 68.218 & 0.059 & 2.502 & 3.866 & 3.256 & 0.96 & -0.072 & 0.098 \\ 
   &  &  & 0.1 & 3.228 & 0.251 & 0.189 & 61.883 & 0.061 & 2.531 & 3.925 & 3.292 & 0.962 & -0.064 & 0.106 \\ 
   &  &  & 0.7 & 4.03 & 0.256 & 0.21 & 69.202 & 0.063 & 3.319 & 4.74 & 4.109 & 0.972 & -0.08 & 0.231 \\ 
   & 50\% & normal & 0 & 30.452 & 0.31 & 0.221 & 35.27 & 0.09 & 29.591 & 31.312 & 30.47 & 0.949 & -0.018 & 0.077 \\ 
   &  &  & 0.1 & 30.462 & 0.307 & 0.221 & 38.968 & 0.09 & 29.61 & 31.314 & 30.475 & 0.923 & -0.013 & 0.087 \\ 
   &  &  & 0.7 & 30.487 & 0.301 & 0.224 & 42.769 & 0.086 & 29.65 & 31.323 & 30.502 & 0.926 & -0.015 & 0.146 \\ 
   &  & skewed & 0 & 3.119 & 0.454 & 0.365 & 20.317 & 0.208 & 1.858 & 4.381 & 3.256 & 0.963 & -0.136 & 0.091 \\ 
   &  &  & 0.1 & 3.151 & 0.475 & 0.388 & 19.111 & 0.227 & 1.833 & 4.469 & 3.292 & 0.977 & -0.142 & 0.1 \\ 
   &  &  & 0.7 & 3.919 & 0.509 & 0.435 & 17.998 & 0.252 & 2.505 & 5.333 & 4.109 & 0.97 & -0.19 & 0.219 \\ 
  MAR & 25\% & normal & 0 & 30.425 & 0.184 & 0.151 & 105.503 & 0.032 & 29.915 & 30.935 & 30.47 & 0.906 & -0.045 & 0.078 \\ 
   &  &  & 0.1 & 30.429 & 0.18 & 0.151 & 114.155 & 0.031 & 29.93 & 30.927 & 30.475 & 0.9 & -0.047 & 0.089 \\ 
   &  &  & 0.7 & 30.444 & 0.181 & 0.159 & 118.405 & 0.031 & 29.94 & 30.948 & 30.502 & 0.896 & -0.057 & 0.148 \\ 
   &  & skewed & 0 & 3.066 & 0.184 & 0.149 & 103.034 & 0.032 & 2.554 & 3.577 & 3.256 & 0.856 & -0.19 & 0.072 \\ 
   &  &  & 0.1 & 3.091 & 0.18 & 0.149 & 108.706 & 0.031 & 2.59 & 3.592 & 3.292 & 0.843 & -0.202 & 0.078 \\ 
   &  &  & 0.7 & 3.77 & 0.183 & 0.162 & 119.094 & 0.032 & 3.261 & 4.278 & 4.109 & 0.708 & -0.34 & 0.149 \\ 
   & 50\% & normal & 0 & 30.403 & 0.284 & 0.207 & 47.457 & 0.079 & 29.615 & 31.191 & 30.47 & 0.851 & -0.067 & 0.083 \\ 
   &  &  & 0.1 & 30.402 & 0.281 & 0.205 & 46.188 & 0.076 & 29.621 & 31.182 & 30.475 & 0.838 & -0.073 & 0.093 \\ 
   &  &  & 0.7 & 30.407 & 0.279 & 0.211 & 48.836 & 0.074 & 29.633 & 31.182 & 30.502 & 0.82 & -0.094 & 0.148 \\ 
   &  & skewed & 0 & 2.904 & 0.327 & 0.234 & 34.697 & 0.103 & 1.996 & 3.811 & 3.256 & 0.866 & -0.352 & 0.064 \\ 
   &  &  & 0.1 & 2.936 & 0.332 & 0.237 & 33.535 & 0.106 & 2.013 & 3.859 & 3.292 & 0.867 & -0.356 & 0.07 \\ 
   &  &  & 0.7 & 3.53 & 0.34 & 0.253 & 33.878 & 0.112 & 2.586 & 4.474 & 4.109 & 0.734 & -0.579 & 0.114 \\ 
   \hline
\end{tabular}
\end{sidewaystable}

\begin{sidewaystable}
\centering
\caption{Method scaled, blend = 0.\label{tab7}}%
\begin{tabular}{lllllllllllllll}
  \hline
mech & mis & dist & cor & qbar & se & t & df & b & 2.5\% & 97.5\% & true & cov & bias & R2 \\ 
  \hline
MCAR & 25\% & normal & 0 & 30.457 & 0.143 & 0.134 & 163.429 & 0.019 & 30.059 & 30.855 & 30.47 & 0.893 & -0.012 & 0.074 \\ 
   &  &  & 0.1 & 30.462 & 0.146 & 0.136 & 157.712 & 0.02 & 30.058 & 30.867 & 30.475 & 0.895 & -0.013 & 0.085 \\ 
   &  &  & 0.7 & 30.496 & 0.146 & 0.144 & 167.413 & 0.02 & 30.091 & 30.901 & 30.502 & 0.887 & -0.006 & 0.144 \\ 
   &  & skewed & 0 & 3.188 & 0.145 & 0.137 & 165.544 & 0.02 & 2.785 & 3.59 & 3.256 & 0.89 & -0.068 & 0.096 \\ 
   &  &  & 0.1 & 3.223 & 0.147 & 0.139 & 159.723 & 0.021 & 2.814 & 3.633 & 3.292 & 0.891 & -0.069 & 0.103 \\ 
   &  &  & 0.7 & 4.011 & 0.147 & 0.157 & 183.983 & 0.021 & 3.603 & 4.419 & 4.109 & 0.86 & -0.098 & 0.223 \\ 
   & 50\% & normal & 0 & 30.497 & 0.202 & 0.156 & 86.643 & 0.039 & 29.936 & 31.059 & 30.47 & 0.824 & 0.028 & 0.069 \\ 
   &  &  & 0.1 & 30.505 & 0.208 & 0.16 & 83.654 & 0.041 & 29.928 & 31.082 & 30.475 & 0.842 & 0.03 & 0.079 \\ 
   &  &  & 0.7 & 30.532 & 0.209 & 0.167 & 85.635 & 0.041 & 29.953 & 31.111 & 30.502 & 0.843 & 0.03 & 0.132 \\ 
   &  & skewed & 0 & 3.122 & 0.206 & 0.159 & 84.192 & 0.04 & 2.55 & 3.694 & 3.256 & 0.832 & -0.134 & 0.09 \\ 
   &  &  & 0.1 & 3.15 & 0.207 & 0.161 & 83.346 & 0.04 & 2.574 & 3.725 & 3.292 & 0.806 & -0.143 & 0.097 \\ 
   &  &  & 0.7 & 3.847 & 0.209 & 0.18 & 97.013 & 0.042 & 3.266 & 4.428 & 4.109 & 0.774 & -0.263 & 0.2 \\ 
  MAR & 25\% & normal & 0 & 30.402 & 0.15 & 0.136 & 153.443 & 0.021 & 29.987 & 30.817 & 30.47 & 0.856 & -0.068 & 0.077 \\ 
   &  &  & 0.1 & 30.403 & 0.15 & 0.138 & 151.516 & 0.021 & 29.985 & 30.82 & 30.475 & 0.859 & -0.072 & 0.088 \\ 
   &  &  & 0.7 & 30.406 & 0.154 & 0.146 & 153.178 & 0.022 & 29.979 & 30.833 & 30.502 & 0.848 & -0.096 & 0.145 \\ 
   &  & skewed & 0 & 3.054 & 0.143 & 0.133 & 162.082 & 0.019 & 2.657 & 3.451 & 3.256 & 0.742 & -0.202 & 0.069 \\ 
   &  &  & 0.1 & 3.084 & 0.145 & 0.135 & 158.785 & 0.02 & 2.681 & 3.486 & 3.292 & 0.755 & -0.209 & 0.074 \\ 
   &  &  & 0.7 & 3.681 & 0.144 & 0.144 & 175.877 & 0.02 & 3.28 & 4.081 & 4.109 & 0.465 & -0.429 & 0.13 \\ 
   & 50\% & normal & 0 & 30.354 & 0.214 & 0.163 & 77.214 & 0.044 & 29.759 & 30.949 & 30.47 & 0.758 & -0.116 & 0.08 \\ 
   &  &  & 0.1 & 30.348 & 0.211 & 0.162 & 82.081 & 0.042 & 29.763 & 30.933 & 30.475 & 0.752 & -0.127 & 0.09 \\ 
   &  &  & 0.7 & 30.318 & 0.214 & 0.171 & 85.563 & 0.043 & 29.725 & 30.911 & 30.502 & 0.739 & -0.184 & 0.141 \\ 
   &  & skewed & 0 & 2.86 & 0.201 & 0.154 & 89.028 & 0.039 & 2.301 & 3.419 & 3.256 & 0.638 & -0.396 & 0.058 \\ 
   &  &  & 0.1 & 2.891 & 0.211 & 0.158 & 76.828 & 0.042 & 2.306 & 3.477 & 3.292 & 0.653 & -0.401 & 0.065 \\ 
   &  &  & 0.7 & 3.438 & 0.204 & 0.163 & 86.894 & 0.039 & 2.871 & 4.005 & 4.109 & 0.395 & -0.672 & 0.1 \\ 
   \hline
\end{tabular}
\end{sidewaystable}

\end{document}